\documentclass[reprent,aps,prb,twocolumn]{revtex4-2}
\usepackage{amsmath,amssymb}
\usepackage[dvipdfmx]{graphicx}
\usepackage{bm}
\usepackage{braket}
\usepackage{color}

\newcounter{num}

\begin{document}
\title{
Emergent cross-product-type spin-orbit coupling under ferroaxial ordering
}

\author{Akane Inda and Satoru Hayami}
\affiliation{
Graduate School of Science, Hokkaido University, Sapporo 060-0810, Japan
}
\begin{abstract}
A ferroic alignment of a time-reversal-even axial vector, which is called the ferroaxial ordered state, is identified by the spontaneous rotational distortion of the lattice structure to break the vertical mirror symmetry. 
We theoretically investigate what happens in the electronic structure once the ferroaxial ordering occurs by analyzing a tetragonal $d$-$p$ cluster model with transition metal oxide in mind. 
By adopting the symmetry-adapted multipole representation, we elucidate the emergent cross-product form of the spin-orbit coupling by the interplay between the atomic spin-orbit coupling and the $d$-$p$ hybridization according to the vertical mirror symmetry breaking, which corresponds to the electronic order parameter of the ferroaxial ordering. 
\end{abstract}

\maketitle
\textit{Introduction.}
In condensed matter physics, various types of spontaneous symmetry breakings owing to electron correlation have been extensively explored.
In particular, the cooperation between the lattice and electron degrees of freedom gives rise to unconventional quantum states of matter, such as orbital ordering by the Jahn-Teller effect~\cite{streltsov2017orbital, PRB_orbital_ordering, PRMat_Orbital_ordering} and noncollinear magnetic ordering by the spin-lattice coupling~\cite{CdCr2O4,CuFeO2,CuAlCr4S8}.

Among them, the ferroaxial ordering has drawn growing interest in recent years since its direct observation in RbFe(MoO$_4$)$_2$~\cite{jin2020natphys_RbFeMoO42, Hayashida2021PRM_ferroaxial_domain} and NiTiO$_3$~\cite{hayashida2020natcom_NiTiO3, Hayashida2021PRM_ferroaxial_domain, yokota2022npj_three-dimensional_imaging, fang2023ferrorotational}. 
These ferroaxial ordered phases are driven by the rotational structural distortion of the ligand sites surrounding the transition metal (TM) sites so as to break the mirror symmetry parallel to the principal axis, whose emergence requires the breakings of neither the spatial inversion ($\mathcal{P}$) symmetry nor the time-reversal ($\mathcal{T}$) symmetry. 
The different $\mathcal{P}$ and $\mathcal{T}$ parities of the ferroaxial ordering from those of the ferromagnetic and ferroelectric orderings indicate the appearance of further intriguing physical properties~\cite{cheong2021permutable, Roy2022prm_spin-current}. 
For example, the ferroaxial ordering exhibits a rotational response of the conjugate physical quantities with the same $\mathcal{P}$ and $\mathcal{T}$ symmetries; an input field along the $x$ ($y$) direction induces the conjugate physical quantities as an output along the $y$ ($-x$) direction~\cite{Hayami2022jpsj_spincurrent}. 
Furthermore, the ferroaxial ordering shows an unconventional optical response such as the electric gyration, which is the optical rotation induced by an external electric field parallel to the principal axis~\cite{hayashida2020natcom_NiTiO3, Hayashida2021PRM_ferroaxial_domain}.

\begin{figure}[th]
  \centering
  \includegraphics[width=\linewidth]{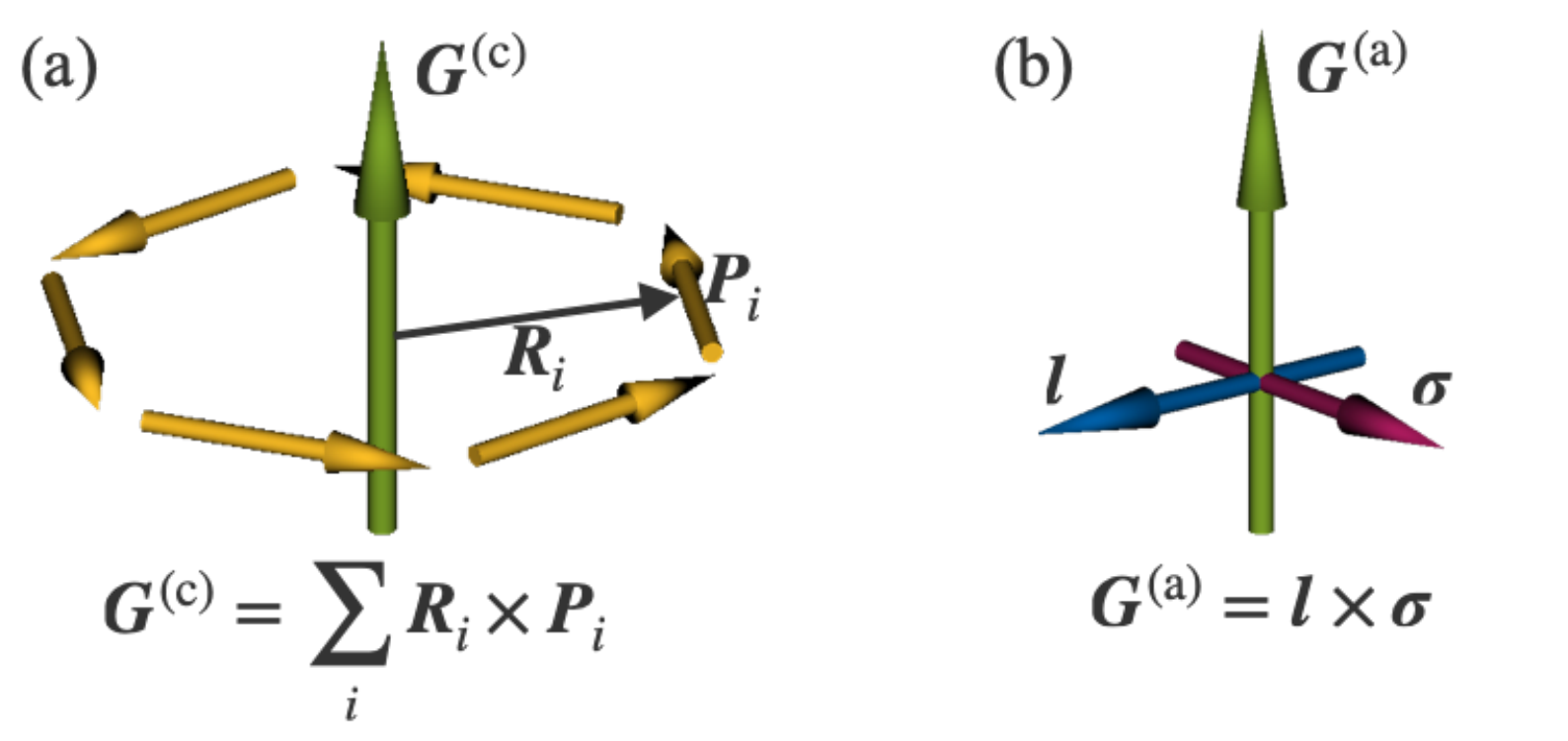}
  \caption{
  \label{f:Gz} 
  Schematic pictures of (a) the cluster ETD $\bm{G}^{\rm (c)}$ (green arrow) and (b) the atomic ETD $\bm{G}^{\rm (a)}$ (green arrow). 
  The cluster ETD is represented as a vortex of electric polarization (yellow arrows), while the atomic ETD is represented as an outer product of orbital angular momentum $\bm{l}$ (blue arrow) and spin angular momentum $\bm{\sigma}$ (purple arrow). 
  }
\end{figure}

From the symmetry viewpoint, the ferroaxial ordered phase is microscopically described as the ferroic alignment of the $\mathcal{T}$-even axial vector.
Since such an axial vector belongs to the identity irreducible representation in 13 out of 32 point groups~\cite{hlinka2016aps_Symmetry_Guide_to_Ferroaxial_Transitions}, there are various candidate ferroaxial materials, such as Cu$_3$Nb$_2$O$_8$~\cite{johnson2011prl_Cu3Nb2O8}, CaMn$_7$O$_{12}$~\cite{johnson2012prl_CaMn7O12}, BaCoSiO$_4$~\cite{xu2022prb_BaCoSiO4}, Ca$_5$Ir$_3$O$_{12}$~\cite{Hasegawa_doi:10.7566/JPSJ.89.054602, hanate2021first, hayami2023cluster, hanate2023space}, K$_2$Zr(PO$_4$)$_2$~\cite{yamagishi2023ferroaxial}, Na$_2$Hf(BO$_3$)$_2$~\cite{nagai2023chemicalSwitching}, Na-superionic conductors~\cite{nagai2023chemical}, MnTiO$_3$~\cite{Sekine_PhysRevMaterials.8.064406}, and 1$T$-TaS$_2$~\cite{Luo_PhysRevLett.127.126401, liu2023electrical}. 
Nevertheless, the effect of the rotational distortion on the electronic structures has been less studied~\cite{Hayami2022jpsj_spincurrent, bhowal2024electric}, and the fundamental issue of how the lattice and electronic degrees of freedom are coupled has not been elucidated so far.

The symmetry-adapted multipole representation reveals that the ferroaxial order parameter is the electric toroidal dipole (ETD)~\cite{hayami2024unified}, which is often defined as the vortex consisting of multiple electric dipoles~\cite{dubovik1975multipole, dubovik1986axial, dubovik1990toroid};
\begin{align}
\label{eq: Gc}
  \bm{G}^{\rm (c)} = \sum_i \bm{R}_i\times \bm{P}_i, 
\end{align}
where $\bm{R}_i$ and $\bm{P}_i$ are the position and electric polarization of site $i$, respectively, as shown in Fig.~\ref{f:Gz}(a). 

Although the quantity in Eq.~(\ref{eq: Gc}) provides an intuitive understanding of the emergence of the ferroaxial ordering in rotationally distorted systems, the choice of the cluster origin is ambiguous in a periodic crystal structure. 
In addition, the description based on Eq.~(\ref{eq: Gc}) is sometimes ill-defined, since the electric polarization is not directly related to the rotational distortion of the ligand sites.

Alternatively, the atomic description independent of the cluster origin is introduced as~\cite{kusunose2020jpsj_completebasis} 
\begin{align}
  \label{eq: Gdef}
  \bm{G}^{\rm (a)} = \bm{l} \times \bm{\sigma},
\end{align}
where $\bm{l}$ and $\bm{\sigma}=2\bm{s}$ are the orbital and spin angular momenta as shown in Fig.~\ref{f:Gz}(b), respectively. 
This corresponds to the emergent cross-product-type spin-orbit coupling (SOC), which is qualitatively different from the conventional relativistic SOC defined by the inner product as $\bm{l}\cdot \bm{\sigma}$. 
In other words, further intriguing spin-orbital-lattice entangled physics is expected under the ferroaxial ordering. 
However, the microscopic origin of such an atomic ETD has not been clarified.
In the present study, we show the microscopic origin of the atomic ETD by analyzing a fundamental $d$-$p$ cluster model. 
We find that the interplay between the relativistic atomic SOC and the $d$-$p$ hybridization caused by the rotational distortion of the ligand sites is a source of the unconventional cross-product-type SOC.

\begin{figure}[t]
  \centering
  \includegraphics[width=\linewidth]{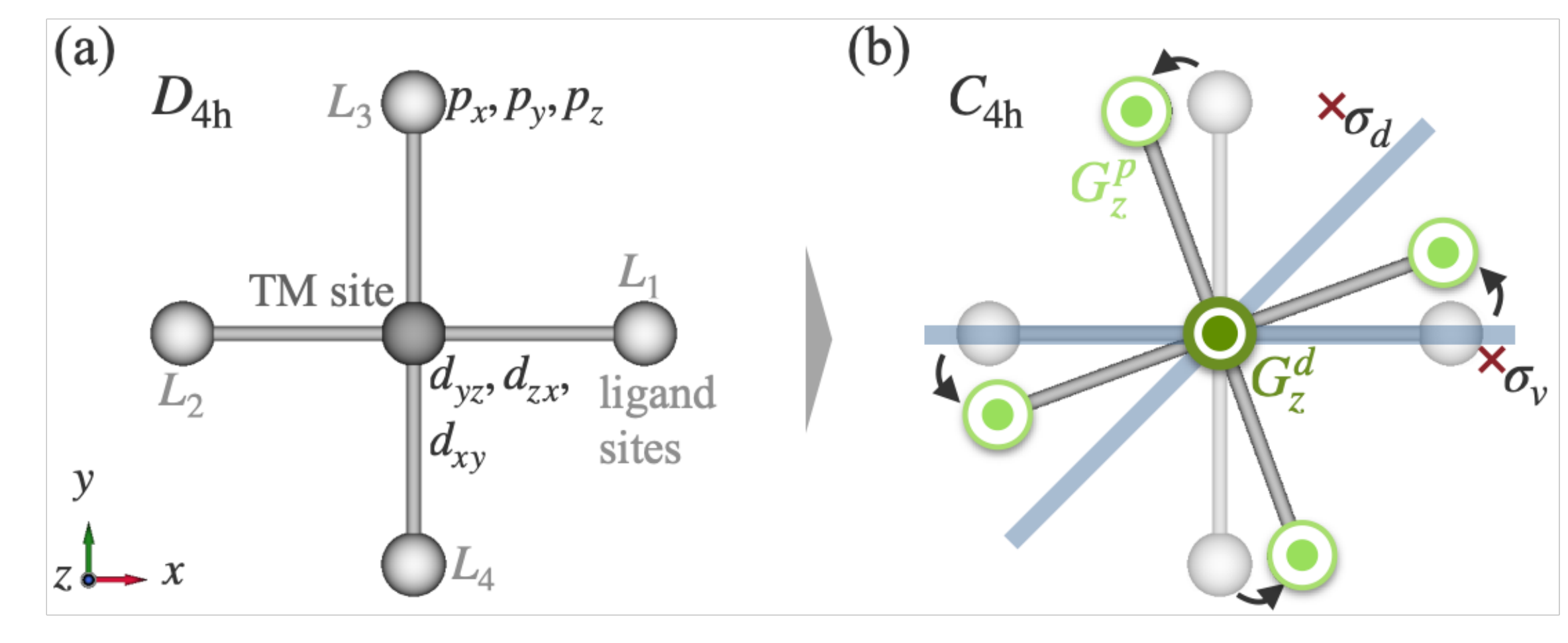}
  \caption{
  \label{f:model}
  Schematic structure of the five-site tetragonal cluster model under the point group (a) $D_{\rm 4h}$ and (b) $C_{\rm 4h}$. 
  The TM sites consist of $d_{yz}, d_{zx}, d_{xy}$ orbitals and the ligand sites ($L_1$--$L_4$) consist of $p_x, p_y, p_z$ orbitals. 
  The $z$-component of the atomic ETD on the TM site ($G^d_z$) and ligand sites ($G^p_z$), which are represented by dark green and light green circles, respectively, are activated according to the vertical mirror symmetry ($\sigma_v$ and $\sigma_d$) breakings.
  }
\end{figure}

\textit{Model.} 
With the TM oxide consisting of the TM surrouded by four ligand oxygens in mind, we consider a fundamental $d$-$p$ model under the tetragonal $D_{\rm 4h}$ symmetry; there are three $d$ orbitals $(d_{yz},d_{zx},d_{xy})$ at the TM site and three $p$ orbitals $(p_x, p_y, p_z)$ at the ligand sites $L_1$--$L_4$ as shown in Fig.~\ref{f:model}(a). 
The model Hamiltonian is given as 
\begin{align}
\label{eq: Hamtotal}
  \mathcal{H} &=\mathcal{H}_{\rm CEF}+\mathcal{H}_{\rm SOC}+
  \mathcal{H}_{dp}.
\end{align}
The first term $\mathcal{H}_{\rm CEF}$ represents the CEF Hamiltonian, where the atomic-energy levels for the $(d_{yz},d_{zx},d_{xy})$ orbitals under the $D_{\rm 4h}$ site symmetry are $(1.7, 1.7, 1.4)$ at the TM site and those for the $(p_x, p_y, p_z)$ orbitals under the $C_{\rm 2v}$ site symmetry are $(0.4, 0.2, 0)$ at the $L_1$ site; the $p_z$ level is taken for the energy origin. 
The second term $\mathcal{H}_{\rm SOC}$ represents the atomic SOC for the $d$ orbitals at the TM site and for the $p$ orbitals at the ligand sites, which are represented by
\begin{align}
  \mathcal{H}_{\rm SOC} &=\lambda_p\sum_k \boldsymbol{l}^p_k \cdot \boldsymbol{s}^p_k
  +\lambda_d \boldsymbol{l}^d \cdot \boldsymbol{s}^d,
\end{align}
where $\bm{l}^d$ and $\bm{s}^d$ ($\bm{l}^p_k$ and $\bm{s}^p_k$ for $k=L_1$--$L_4$) are the orbital and spin angular momentum operators at the TM (ligand) site, respectively. 
The third term represents the $d$-$p$ hybridization between the TM and ligand sites, which is given by
\begin{align}
  \label{eq: Hdp}
  \mathcal{H}_{dp}&= 
  \sum_{k=L_1}^{L_4}\sum_{\alpha\beta\sigma} t_{\alpha\beta}^{{\rm A}_{1g}} c_{\alpha\sigma}^\dagger c_{\beta\sigma}+({\rm h.c.}),
\end{align}
where $c_{\alpha\sigma}^\dagger$ ($c_{\alpha\sigma}$) is the creation (annihilation) operator; $\alpha$ represents the $(d_{yz}, d_{zx}, d_{xy})$ orbitals, $\beta$ represents a set of the $(p_{x}, p_y, p_z)$ orbitals and the ligand sites $L_1$--$L_4$, and $\sigma$ represents the up and down spins. 
By using the Slater-Koster parameters, there are two symmetry-allowed hybridization parameters: One is the hybridization between $d_{xy}$ and $p_{y}$, $t^{{\rm A}_{1g}}_{y,xy}$, and the other is that between $d_{zx}$ and $p_z$ orbitals, $t^{{\rm A}_{1g}}_{z,zx}$, for the $L_1$ site. 
The hybridization parameters $t_{\alpha\beta}^{{\rm A}_{1g}}$ for $k = L_2$--$L_4$ are obtained by the fourfold rotation around the TM site.

When the rotational structural distortion by the ligand sites to break the vertical mirror symmetry occurs as shown in Fig.~\ref{f:model}(b), the point group symmetry reduces to $C_{\rm 4h}$ and the ${\rm A}_{2g}$ representation under $D_{\rm 4h}$ belong to the identity irreducible representation ${\rm A}_g$. 
Since the cross-product-type SOC in Eq.~(\ref{eq: Gdef}) also belongs to the ${\rm A}_{2g}$ representation under $D_{\rm 4h}$, one can expect nonzero expectation values of the atomic ETD under $C_{\rm 4h}$. 
To understand the microscopic origin of such an ETD, we focus on the modulation of the $d$-$p$ hybridizations induced by the rotational distortion. 
Reflecting the lack of vertical mirror symmetry in Fig.~\ref{f:model}(b), additional two hybridization parameters are symmetry-allowed in Eq.~(\ref{eq: Hdp}), which correspond to the hybridization between $d_{xy}$ and $p_x$ orbitals, $ t_{x,xy}^{{\rm A}_{2g}}$, and that between $d_{yz}$ and $p_z$ orbitals, $ t_{z,yz}^{{\rm A}_{2g}}$, in the case of the $L_1$ site.

\textit{Numerical results.}
\begin{figure}[t!]
  \centering
  \includegraphics[width=\linewidth]{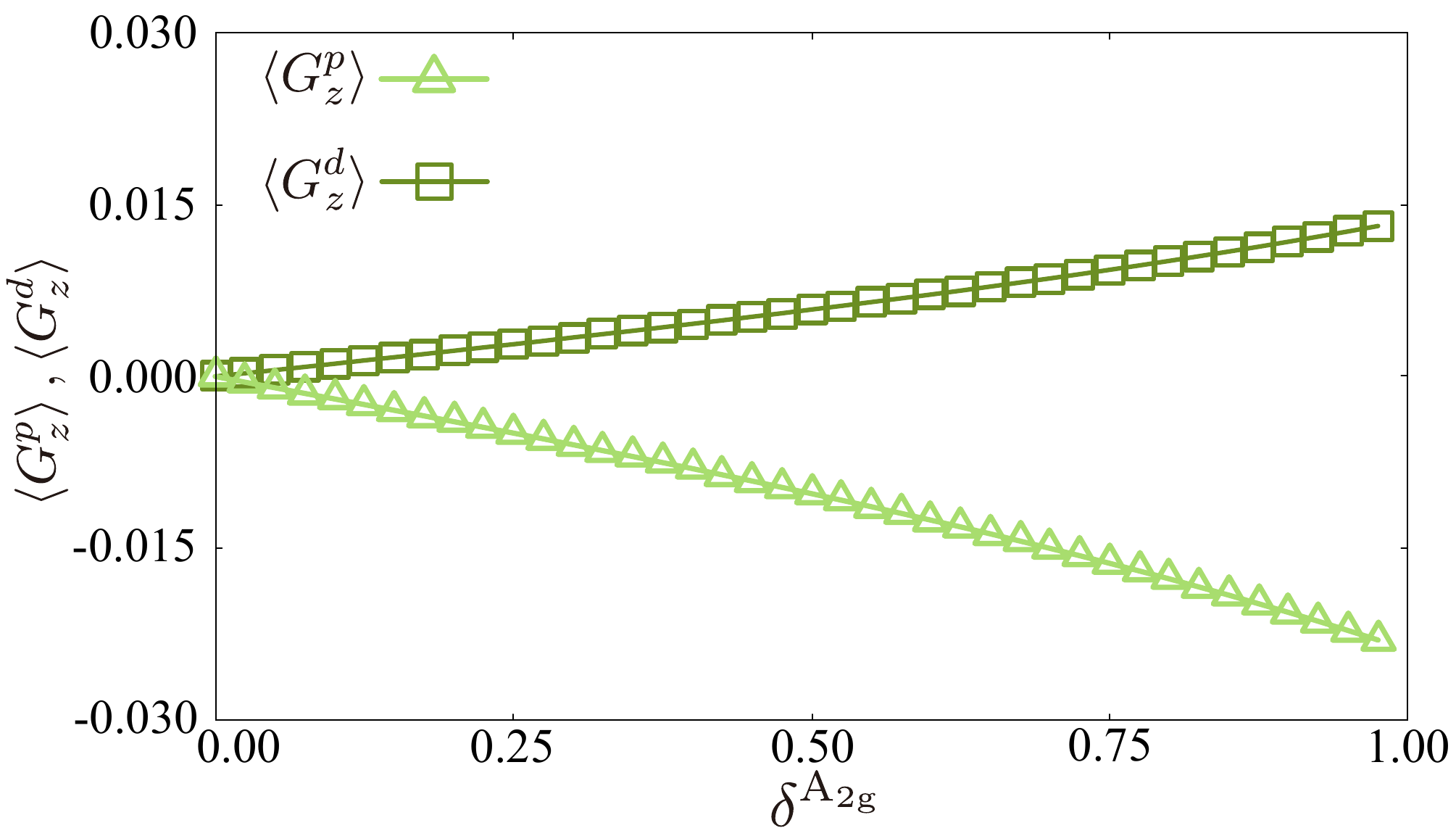}
  \caption{
  \label{f:distortion_dependence} 
  Expectation values of the ETD for the $p$ orbital $\braket{G^{p}_z}$ and that for the $d$ orbital $\braket{G_z^d}$ against the hybridization originating from the rotational distortion, $\delta^{{\rm A}_{2g}}$, for the filling per site $n=28/5$, $t_{x,xy}^{{\rm A}_{2g}} = 0.4 \delta^{{\rm A}_{2g}}$, $t_{z,yz}^{{\rm A}_{2g}} = 0.2 \delta^{{\rm A}_{2g}}$, $t_{y,xy}^{{\rm A}_{1g}}= 0.6$, $t_{z,zx}^{{\rm A}_{1g}}= 1.4$, $\lambda_p = 0.05$, and $\lambda_{d} = 0.07$.
  }
\end{figure} 
Figure~\ref{f:distortion_dependence} represents the behavior of the atomic ETD when the hybridization parameters, $t_{x,xy}^{{\rm A}_{2g}}$ and $t_{z,yz}^{{\rm A}_{2g}}$, are varied.
Here, we evaluate the expectation values of the ETD operators for the $p$ and $d$ orbitals, which are defined by $G^p_z=\sum_k (\boldsymbol{l}_k^p \times \boldsymbol{\sigma}_k^p)_z$ and $G^d_z=(\boldsymbol{l}^d \times \boldsymbol{\sigma}^d)_z$~\cite{Kusunose_PhysRevB.107.195118}. 
In addition, we parametrize the modulation of the hybridization as $\delta^{{\rm A}_{2g}}$ with $t_{x,xy}^{{\rm A}_{2g}} = 0.4 \delta^{{\rm A}_{2g}}$ and $t_{z,yz}^{{\rm A}_{2g}} = 0.2 \delta^{{\rm A}_{2g}}$. 
The other model parameters are set as $t_{y,xy}^{{\rm A}_{1g}}= 0.6$, $t_{z,zx}^{{\rm A}_{1g}}= 1.4$, $\lambda_p = 0.05$, and $\lambda_{d}= 0.07$ for the filling per site $n=28/5$. 
The result shows that both $\braket{G^p_z}$ and $\braket{G^d_z}$ become finite for nonzero $\delta^{{\rm A}_{2g}}$, which indicates that the cross-product-type SOC emerges according to the vertical mirror symmetry breaking.

\begin{figure}[t!]
  \centering
  \includegraphics[width=\linewidth]{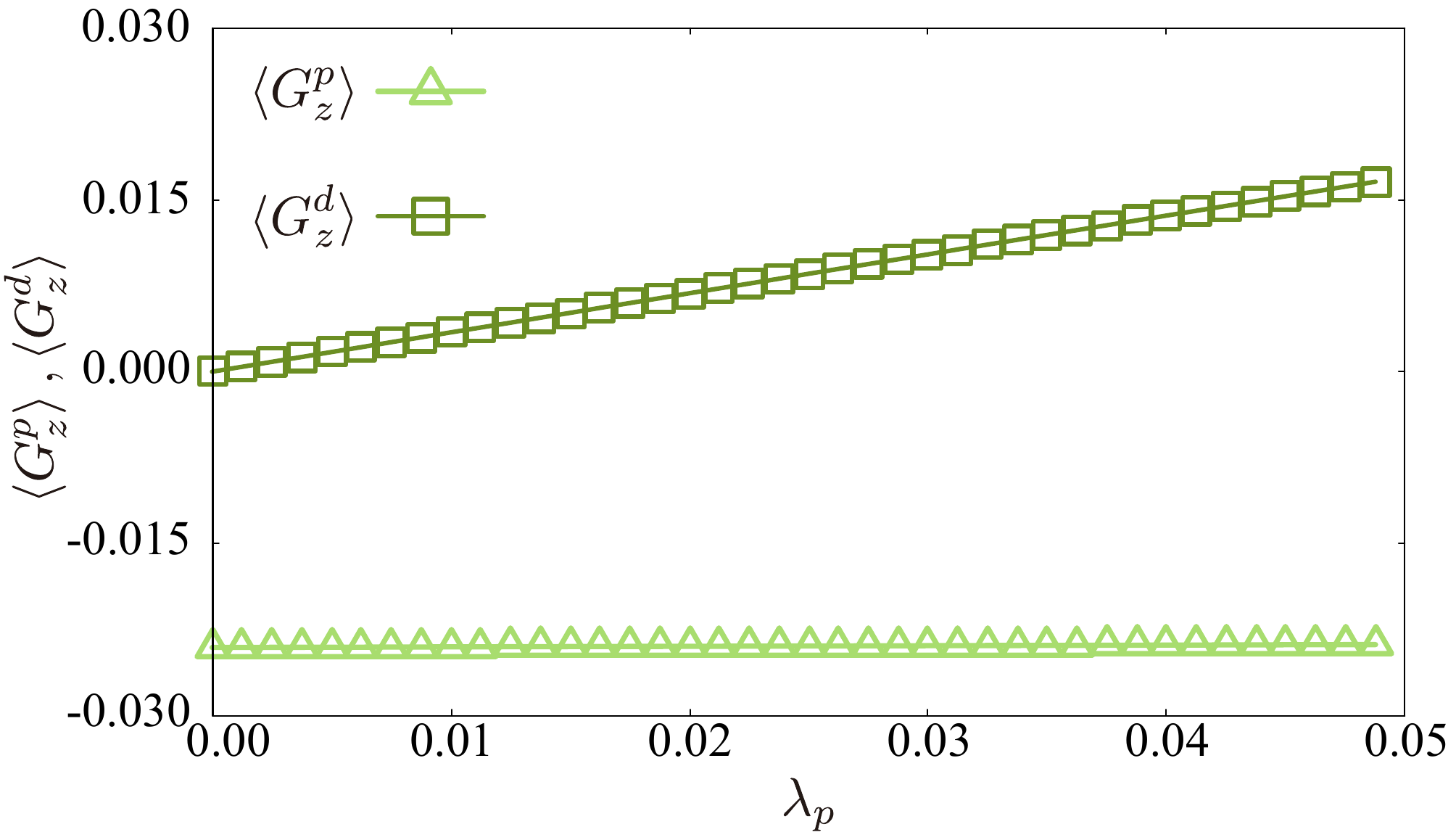}
  \caption{
  \label{f:lambda_p_dependence} 
  $\lambda_p$ dependence of $\braket{G^{p}_z}$ and $\braket{G_z^{d}}$, for the filling per site $n=28/5$, $t_{x,xy}^{{\rm A}_{2g}} = 0.4$, $t_{z,yz}^{{\rm A}_{2g}} = 0.2$, $t_{y,xy}^{{\rm A}_{1g}}= 0.6$, $t_{z,zx}^{{\rm A}_{1g}}= 1.4$, and $\lambda_d = 0.07$. 
    }
\end{figure}

Let us show the important microscopic degrees of freedom to induce the cross-product-type SOC. 
The one is the atomic SOC. 
The SOC for the $p$ orbital is necessary for nonzero $\braket{G^d_z}$, as demonstrated by its $\lambda_{p}$ dependence in Fig.~\ref{f:lambda_p_dependence} by fixing $t_{x,xy}^{{\rm A}_{2g}} = 0.4$ and $t_{z,yz}^{{\rm A}_{2g}} = 0.2$ while keeping the other parameters; $\braket{G^d_z}$ vanishes for $\lambda_p=0$, while $\braket{G^p_z}$ is less sensitive to $\lambda_p$. 
The opposite tendency occurs when $\lambda_d=0$ instead of $\lambda_p$; $\braket{G^p_z}$ vanishes and $\braket{G^d_z}$ takes nonzero values. 
Thus, $\braket{G^d_z}$ and $\braket{G^p_z}$ vanish in the absence of the SOC for the other orbitals. 
The other is the three $d$ and $p$ orbital degrees of freedom. 
Among them, $p_z$ and $d_{xy}$ orbitals are necessary since $\braket{G^d_z}$ ($\braket{G^p_z}$) vanishes when $p_z$ ($d_{xy}$) orbital is neglected, where the ETD operator $G_z^p$ ($G_z^d$) is no longer defined in this case.
In addition, the other four orbitals are also necessary to define the ETD operator in the $p$ and $d$ orbital space so that the orbital angular momentum $\bm{l}$ can be activated. 
Therefore, the $d$-$p$ model consisting of six orbitals in Eq.~(\ref{eq: Hamtotal}) is a minimal model to capture the origin of the cross-product-type SOC.

\begin{figure}[thb]
  \centering
  \includegraphics[width=\linewidth]{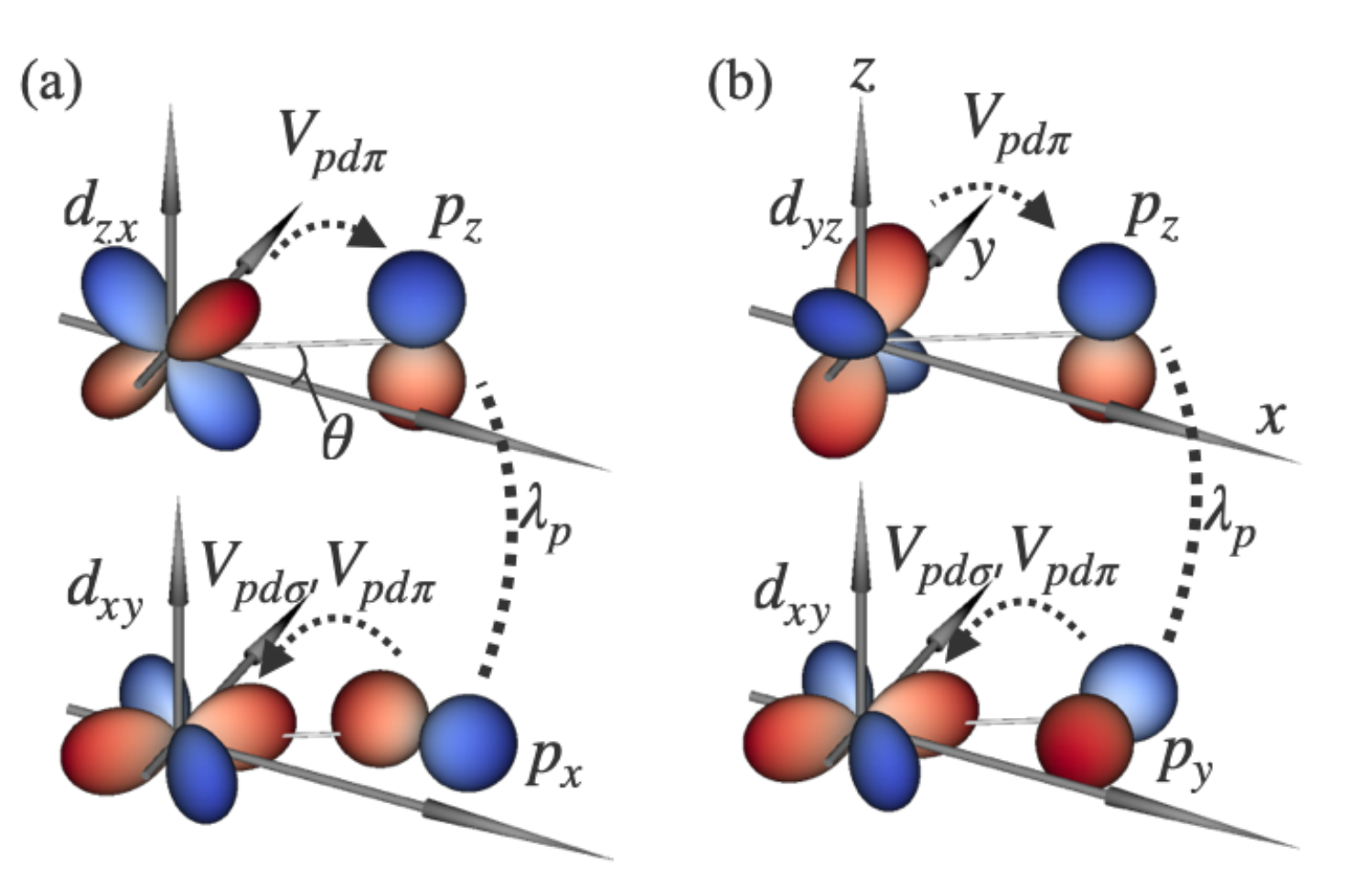}
  \caption{
  \label{f:path}
  Two effective hybridization paths between the TM site and the ligand $L_1$ site to induce the atomic ETD (cross-product-type SOC) at the TM site, which are represented by (a) $\lambda_p t_{z,zx}^{{\rm A}_{1g}}t_{x,xy}^{{\rm A}_{2g}}$ and (b) $\lambda_p t_{y,xy}^{{\rm A}_{1g}}t^{{\rm A}_{2g}}_{z,yz}$. 
  }
\end{figure}

\textit{Perturbation analysis.}
To further get a deep insight into the microscopic origin of the cross-product-type SOC, we perform the perturbative analysis; we focus on the important ingredients for obtaining nonzero $\braket{G^d_z}$. 
For this purpose, we simplify the $d$-$p$ model in Eq.~(\ref{eq: Hamtotal}) by neglecting the atomic SOC for the $d$ orbitals and supposing the same CEF level for the three $d$ and $p$ orbitals as $\Delta$ and $0$, respectively, since these factors are not important in the following results. 
The simplified Hamiltonian is given by 
\begin{align}
  \mathcal{H}_{\rm tot} &=\mathcal{H}_0 +\mathcal{H}_{dp},\\
  \mathcal{H}_0 &=\Delta \sum_{\alpha, \sigma}
  c_{\alpha \sigma}^\dagger c_{\alpha \sigma}
  +\lambda_p\sum_k  \boldsymbol{l}^p_k \cdot \boldsymbol{s}^p_k.
\end{align}
Using the perturbative wave functions within the second order of $\mathcal{H}_{dp}$, the effective hybridization paths to couple with the atomic ETD for the $d$ orbital along the $L_1$ site are derived as
\begin{align}
\label{eq: perturbative}
  {\rm Tr}[\mathcal{H}_{L_1}^{(2)} G_z^{d}]&=\frac{4\lambda_p (t^{{\rm A}_{1g}}_{z,zx}t^{{\rm A}_{2g}}_{x,xy}-t^{{\rm A}_{1g}}_{y,xy}t^{{\rm A}_{2g}}_{z,yz})}{(\Delta-\lambda_p) (\lambda_p+2 \Delta)},
\end{align}
where $\mathcal{H}_{L_1}^{(2)}$ stands for the contribution from the $L_1$ site in  the effective Hamiltonian obtained by the second-order perturbation in terms of $\mathcal{H}_{dp}$, whose general form is given by
\begin{align}  
  \mathcal{H}^{(2)} &= \sum_{\alpha,\alpha'\tilde{\beta} \sigma''} \frac{
  \langle d_{\alpha,\sigma}^{(0)}|\mathcal{H}_{dp} |p_{\tilde{\beta},\sigma''}^{(0)}\rangle
  \langle p_{\tilde{\beta},\sigma''}^{(0)}|\mathcal{H}_{dp}|d_{\alpha',\sigma'}^{(0)}\rangle 
  }{E_{\alpha}^{(0)}-E_{\tilde{\beta},\sigma''}^{(0)}}. \label{eq.second_pertu}
\end{align}
Here, $\ket{p^{(0)}_{\tilde{\beta},\sigma''}}$ and $\ket{d^{(0)}_{\alpha', \sigma'}}$ ($E_{\tilde{\beta},\sigma''}^{(0)}$ and $E_{\alpha}^{(0)}$) represent the eigenvectors (energy levels) of $\mathcal{H}_0$ for spin $\sigma', \sigma'' = \uparrow,\downarrow$, respectively; the subscripts $\alpha, \alpha'$ ($\tilde{\beta}$) stand for the index for $yz$, $zx$, and $xy$  ($x$,$y$, and $z$) orbitals. 
$E^{(2)}_{\alpha}$ denotes the second-order energy correction and $\bm{1}$ denotes the $6\times 6$ identity matrix.

By taking the limit of the large energy difference between the $d$ and $p$ orbitals ($\Delta \gg \lambda_p$), the expression in Eq.~(\ref{eq: perturbative}) turns into
\begin{align}  
  {\rm Tr}[\mathcal{H}_{L_1}^{(2)} G_z^{d}]
  &\simeq
  \frac{2\lambda_p }{\Delta^2}
  (t_{z,zx}^{{\rm A}_{1g}}t_{x,xy}^{{\rm A}_{2g}}-t_{y,xy}^{{\rm A}_{1g}}t^{{\rm A}_{2g}}_{z,yz})
  \label{eq.effective_path}\\
  &\propto
  \lambda_p \sin 4\theta
  (\sqrt{3}V_{pd\sigma}-2V_{pd\pi})V_{pd\pi}
  \label{eq.Slater_Koster}.
\end{align}
In the last line, we introduce the Slater-Koster parameters, $V_{pd\sigma}$ and $V_{pd\pi}$, and $\theta$ corresponding to the rotational angle measured from the $x$-axis as shown in Fig.~\ref{f:path}(a).

The expression in Eq.~(\ref{eq.effective_path}) reveals the important hybridization paths as well as the atomic SOC for obtaining the atomic ETD (cross-product-type SOC); the modulations of the hybridization owing to the rotational distortion, $t_{x,xy}^{{\rm A}_{2g}}$ and $t^{{\rm A}_{2g}}_{z,yz}$, contribute to the atomic ETD in collaboration with the ${\rm A}_{1g}$-type hybridizations, $t_{z,zx}^{{\rm A}_{1g}}$ and $t_{y,xy}^{{\rm A}_{1g}}$, for the $L_1$ site, respectively. 
In addition, one finds that such a coupling between ${\rm A}_{1g}$-type and ${\rm A}_{2g}$-type hybridizations is enabled by the atomic SOC for the $p$ orbital $\lambda_p$. 
In a similar manner, the conditions of inducing the atomic ETD for the $p$ orbital is obtained by replacing $\lambda_p$ with $-\lambda_d$ in Eq.~(\ref{eq: perturbative}). 
This perturbative result is consistent with the numerical result discussed above. 
Moreover, as shown in Figs.~\ref{f:path}(a) and \ref{f:path}(b), two effective hybridization paths include $p_z$ orbital.
This is why the $p_z$ orbital is necessary for inducing the atomic ETD at the TM site.

From Eq.~(\ref{eq.Slater_Koster}), the effective hybridization vanishes for $\theta = \pi n/4 (n \in \mathbb{Z})$.
This is understood from the fact that the $x$ direction is equivalent to the $y$ direction for $\theta=\pi n/4$, which means the recovery of the vertical mirror symmetry.
In other words, the point group symmetry remains $D_{\rm 4h}$ for $\theta = \pi n/4$. 
This result also indicates that the lack of vertical mirror symmetry is essential for inducing the atomic ETD, which is consistent with the symmetry consideration.

\textit{Magnetic response.}
\begin{figure}[t!]
  \centering
  \includegraphics[width=\linewidth]{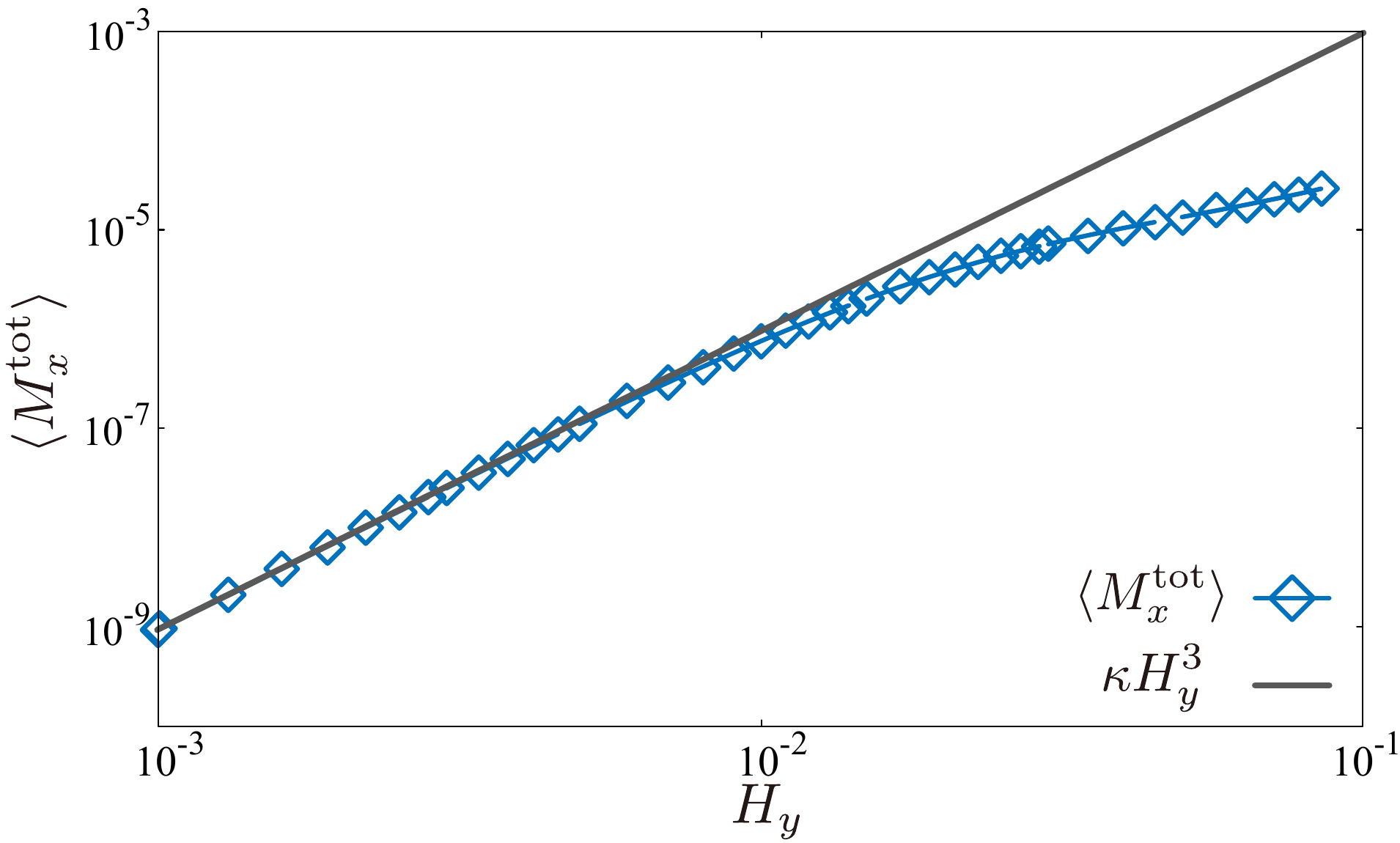}
  \caption{
  \label{f:trasverse_m}
  $H_y$ dependence of $\braket{M_x^{\rm tot}}$ at $t_{x,xy}^{{\rm A}_{2g}} = 0.4$, $t_{z,yz}^{{\rm A}_{2g}} = 0.2$, $t_{y,xy}^{{\rm A}_{1g}}= 0.6$, $t_{z,zx}^{{\rm A}_{1g}}= 1.4$, $\lambda_p = 0.05$ $\lambda_d = 0.07$, and the filling per site $n=28/5$. 
  The gray line shows the fitting line by $\kappa H_y^3$ with $\kappa= 0.948377$.
  }
\end{figure}
Once the ferroaxial ordering occurs, various physical phenomena are expected to emerge, such as the electric gyration~\cite{hayashida2020natcom_NiTiO3, Hayashida2021PRM_ferroaxial_domain}, spin current generation~\cite{Roy2022prm_spin-current, Hayami2022jpsj_spincurrent}, antisymmetric thermopolarization~\cite{nasu2022prb_thermopolarization}, third-order transverse magnetization~\cite{inda2023jpsj}, second-order nonlinear magnetostriction~\cite{kirikoshi2023rotational}, and magnetic instability~\cite{inda2024PRB}. 
Among them, we here evaluate the third-order transverse magnetization under the ferroaxial ordering. 
The effect of an external magnetic field is introduced as the Zeeman coupling effect given by 
\begin{align}
  \mathcal{H}_{\rm ex} &= - \mu_{\rm B}\bm{M}^{\rm tot} \cdot \bm{H},
\end{align}
where $\bm{M}^{\rm tot}= \bm{l}^d + 2\bm{s}^d +\sum_k ( \bm{l}_k^p + 2\bm{s}_k^p)$ is the uniform magnetization and the Bohr magneton $\mu_{\rm B}$ is set as unity. 
When the ferroaxial state occurs, the third-order transverse magnetization as $M^{\rm tot}_x=\chi_{xyyy}H_y^3$ and $M^{\rm tot}_y=-\chi_{yxxx}H_x^3$ is expected from the symmetry. 
The numerical evaluation indicates that the transverse magnetization $\braket{M^{\rm tot}_x}$ is induced in the order of $H_y^3$, as shown in Fig.~\ref{f:trasverse_m}, as expected from the symmetry analysis; model parameters are set as the filling per site $n=28/5$, 
$t_{x,xy}^{{\rm A}_{2g}} = 0.4$, $t_{z,yz}^{\rm A_{2g}} = 0.2$, $t_{y,xy}^{{\rm A}_{1g}}= 0.6$, $t_{z,zx}^{{\rm A}_{1g}}= 1.4$, $\lambda_p = 0.05$, and $\lambda_d = 0.07$. 
We also confirm the relation of $M_x/H_y^3 = -M_y/H_x^3$ for the small magnetic field. 
Such a transverse magnetization becomes a macroscopic experimental signal of the ferroaxial ordering.

\textit{Summary.} 
We have proposed the emergent cross-product-type SOC, which corresponds to the atomic ETD, under the ferroaxial ordering with the rotational structural distortion. 
By analyzing the $d$-$p$ tetragonal cluster model, we have shown that the synergy between the atomic SOC and the $d$-$p$ hybridization arising from the breaking of the vertical mirror symmetries induces the atomic ETD, which can become the origin of the intriguing responses of the conjugate physical quantities as a consequence of the spin-orbital-lattice entanglement.

\textit{Acknowledgments.}
A. I. acknowledges Yi Fan for valuable comments.
This research was supported by JSPS KAKENHI Grants Numbers JP21H01037, JP22H00101, JP22H01183, JP23H04869, JP23K03288, JP23K20827, by JST FOREST (JPMJFR2366), and by JST CREST (JPMJCR23O4).

\bibliographystyle{apsrev}
\bibliography{main}
\end{document}